\documentclass[%
 aip,
 amsmath,amssymb,
 reprint,%
]{revtex4-1}

\usepackage{graphicx}
\usepackage{dcolumn}
\usepackage{bm}
\usepackage{appendix}
\usepackage[utf8]{inputenc}
\usepackage[T1]{fontenc}
\usepackage{mathptmx}
\usepackage{tablefootnote}
\usepackage{longtable}
\newcommand*{\thead}[1]{%
\multicolumn{1}{c}{\bfseries\begin{tabular}{@{}c@{}}#1\end{tabular}}}
\usepackage{color}
\usepackage[table]{xcolor}

\newcommand{\bra}[1]{\ensuremath{\left\langle#1\right|}}
\newcommand{\ket}[1]{\ensuremath{\left|#1\right\rangle}}
\newcommand{\matrixel}[3]{\ensuremath{\langle #1 \left|#2\right| #3 \rangle}}

\begin{document}

\preprint{AIP/123-QED}

\title[]{Resource Efficient Chemistry on Quantum Computers with the Variational Quantum Eigensolver and The Double Unitary Coupled-Cluster approach}

\author{M. Metcalf}
 \email{mmetcalf@lbl.gov}
 \homepage{https://crd.lbl.gov/departments/\\computational-science/ccmc/staff/postdoctoral-fellows/mekena-metcalf/}
 \affiliation{Lawrence Berkeley National Laboratory, 1 Cyclotron Rd, Berkeley, CA 94720}
 \author{N.P.~Bauman}%
 \email{nicholas.bauman@pnnl.gov}
\affiliation{ 
Pacific Northwest National Laboratory, Richland, WA 99352
}%
\author{K.~Kowalski}%
 \email{karol.kowalski@pnnl.gov}
\affiliation{ 
Pacific Northwest National Laboratory, Richland, WA 99352}
\author{W.~A.~de Jong}%
 \email{wadejong@lbl.gov}
\affiliation{Lawrence Berkeley National Laboratory, 1 Cyclotron Rd, Berkeley, CA 94720}%


\date{\today}

\begin{abstract}
Applications of quantum simulation algorithms to obtain electronic energies of molecules on noisy intermediate-scale quantum (NISQ) devices require careful consideration of resources describing the complex electron correlation effects. In modeling second-quantized problems, the biggest challenge confronted is that the number of qubits scales linearly with the size of molecular basis. This poses a significant limitation on the size of the basis sets and the number of correlated electrons included in quantum simulations of chemical processes. 
To address this issue and to enable more realistic simulations on NISQ computers, we employ the double unitary coupled-cluster (DUCC) method to effectively downfold correlation effects into the reduced-size orbital space, commonly referred to as the active space. Using downfolding techniques, we demonstrate that properly constructed effective Hamiltonians can capture the effect of the whole orbital space in small-size active spaces. Combining the downfolding pre-processing technique with the Variational Quantum Eigensolver, we solve for the ground-state energy of $\text{H}_2$ and $\text{Li}_2$ in the cc-pVTZ basis using the DUCC-reduced active spaces. We compare these results to full configuration-interaction and high-level coupled-cluster reference calculations.
\end{abstract}

\maketitle

\begin{figure*}[t]
	\includegraphics[width = 6in]{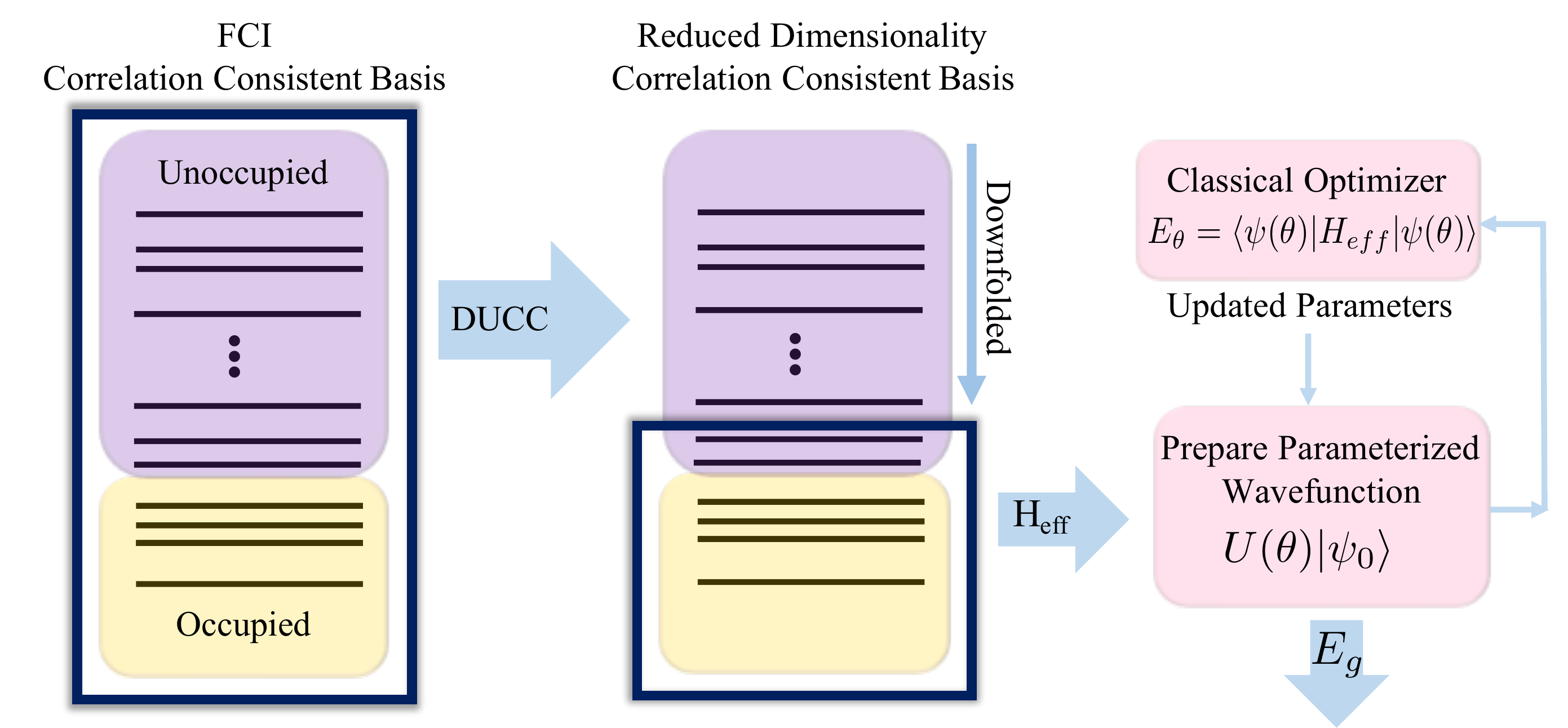}
	\caption{VQE-DUCC Algorithm for discrete molecular Hamiltonians.}
	\label{fig:Illustration}
\end{figure*}
\section{Introduction}

Quantum computing is an exciting prospect for simulating quantum systems on quantum devices~\cite{ZapataRev2019,McArdleReview2019, WhitfieldRev2011}. 
This is a natural consequence of   the fact that 
fermionic creation and annihilation operators along with 
second-quantized molecular Hamiltonians used in theoretical and computational chemistry  can be mapped to qubits.\cite{SeeleyBK2012,BKSF-2018} 
Chemistry applications using quantum computers will likely yield the most benefit for molecules characterized by the presence of strong correlations, where perturbative techniques fail due to convergence problems and consequently are not capable to provide sufficient level of accuracy in calculated ground-state energies. 
Traditional active-space  diagonalization techniques 
such as multi-configurational self-consistent field methods (MCSCF) 
\cite{} or density matrix renormalization group approaches (DMRG) 
\cite{white1992density,schollwock2005density,chan2011density}
can capture only the so-called  static correlation, leaving an important class of dynamical correlation effects unaccounted. Dynamical correlation can be captured through full configuration-interaction (FCI). 
From the perspective of quantum computing, a major limitation to solving relevant chemistry problems with  MCSCF or FCI is the availability of reliable qubits.\cite{MSR-Femoco2017} Near-term quantum computers contain 10's to 100's of noisy qubits, and only limited numbers of operations can be performed. This begs the question, how can quantum chemistry's Hamiltonians be defined and solved within the current and near-term resource limitations and still capture the accuracy of the full configuration-interaction (FCI) solution in the complete basis? 
One approach is to {\it downfold} the higher unoccupied orbitals such that the correlation from the these orbitals is captured in a reduced active space. As discussed in Refs. \cite{bauman2019downfolding,bauman2019quantum}, these formalisms provide a new class of  algorithms for dimensionality-reduction of many-body Hamiltonians based on the concept of an active space.

The correlation effects in strongly correlated molecular systems has been intensively studied over the last few decades.\cite{mrcclyakh}
For example, there are several families of multi-reference perturbative (see Refs.~\onlinecite{finley1995applications,finley1996convergence,chaudhuri2005comparison} and 
references therein) and multi-reference coupled-cluster methods 
\cite{mukherjee1977applications,pal1988molecular,sinha1989eigenvalue,kaldor1991fock,meissner1998fock,musial2008multireference,jezmonk,meissner1,pylypov1,pylypov2,ims1,gms1,mahapatra1,mahapatra2,evangelista1,bwpittner1}
that provide a many-body form of the effective Hamiltonians in the active space.
In our studies, we follow a different strategy based on recent studies of 
the sub-system embedding sub-algebras coupled-cluster formalism (SES-CC) 
\cite{safkk}, where effective Hamiltonians are natural 
consequences of the CC wave function parametrization. From this point of view, 
the SES-CC and closely related double unitary coupled-cluster (DUCC) formalisms 
can be viewed as natural system renormalization procedures.

The DUCC operator leads to a rigorous mathematical algorithm for integrating out fermionic degrees of freedom not included in the active space. In applications discussed in Ref. \onlinecite{bauman2019downfolding}, the choice of the active spaces was driven by energy criteria, which led to the decoupling of two sets of degrees of freedom associated with  low- and high-energy components that define the electronic wave function of interest.
In computational chemistry terms, these two subsets can be identified with static and dynamical correlation effects. However, other  scenarios were envisioned 
for the DUCC formalism where different types of effects -- for example, short- vs long-range correlations effects—are decoupled using an appropriate form of the local DUCC ansatz or an adequate definition of the active space.
The resulting second-quantized downfolded (or effective) Hamiltonians open up the possibility of performing quantum simulations for larger molecular orbital spaces using limited computational resources. 
Recently, these ideas have also been extended to the time-domain
paving the road for a  imaginary-time evolution of downfolded Hamiltonians.\cite{kowalski2020sub}
Simple approximations have been developed and implemented to construct downfolded Hamiltonians, which have been integrated with the Quantum Phase Estimator (QPE) algorithm.\cite{luis1996optimum,
cleve1998quantum,berry2007efficient,childs2010relationship,wecker2015progress,haner2016high,poulin2017fast} 
Preliminary results indicate that this strategy can efficiently encapsulate dynamical correlation in low-rank downfolded Hamiltonians.

We extend the DUCC formalism to variational quantum algorithms~\cite{McClean2016}. These hybrid quantum-classical algorithms prepare trial wavefunctions on a quantum computer with the energy evaluation coupled to a classical optimizer. The variational principal ensures the global minimum yielded by the classical optimizer is the ground-state energy, provided the wave-function ansatz is sufficient in capturing correct the ground state wave function. Variational algorithms require considerably fewer operations (have shallower circuit or gate depth) than algorithms like QPE. Algorithmic tools like dimensionality reduction using DUCC coupled with shallow circuits from chemistry inspired ansatzes are expected to provide a possible pathway to obtain accurate chemistry solutions  with large basis sets on near-term quantum computing hardware. To validate these expectations we benchmark the performance of VQE-DUCC on chemically-relevant molecules. 

This paper is organized as follows: In Section II we introduce the quantum simulation methods employed for chemistry Hamiltonians and the Variational Quantum Eigensolver (VQE) algorithm.
\cite{peruzzo2014variational,McClean2016,openfermion,fontalvo2017strategies,
PhysRevA.95.020501,Kandala2017,kandala2018extending,Colless2018}
In Section III the DUCC formalism is discussed in detail. Section IV presents the performance of the VQE-DUCC algorithms in practice using quantum circuit simulation.

\section{Quantum Simulation of Molecular Hamiltonians}

Properties of a molecule are quantified using the electronic structure Hamiltonian under the Born-Oppenheimer approximation describing electron motion in a field of nuclei
\begin{equation}
    H = -\frac{1}{2}\sum_{i=1}^N \nabla_i^2 - \sum_{i=1}^N\sum_{A=1}^M \frac{Z_A}{r_{iA}} + \sum_{i=1}^N\sum_{j>i}^N \frac{1}{r_{ij}}
\end{equation}
for $N$ electrons and $M$ fixed, point-charge nuclei in continuous space. A central objective is to determine the energy eigenstates of the Hamiltonian in a particular basis $\{\phi({\bf x}_i)\}$ by solving the Schr\"odinger equation
\begin{equation}
H\ket{\Psi}=E\ket{\Psi}
\end{equation}
where $\ket{\Psi}$ represents ground state electronic wave function. 
An alternative representation of the electronic Hamiltonian, known as second quantization, is formed using electron creation $a_i^\dagger \ket{0} = \ket{1}$ and annihilation $a_i\ket{1} = \ket{0}$ field operators with fermion occupation defined as $\ket{1}$($\ket{0}$) for occupied (unoccupied) orbitals. The Hamiltonian is the recast giving

\begin{subequations}
    \begin{equation}
        H = \sum_{ij}h_{ij}a_i^\dagger a_j + \frac{1}{2}\sum_{ijkl}
        h_{ijkl}
        a_i^\dagger a_k^\dagger a_l a_j,
    \end{equation}
    \begin{equation}
        h_{ij} = \int d{\bf x} \phi_i^*({\bf x})\left(\frac{\nabla}{2} - \sum_A^M \frac{Z_A}{r_{iA}}\right)\phi_j({\bf x}),
    \end{equation}
    \begin{equation}
        h_{ijkl} = \int d{\bf x}_1 d{\bf x}_2 \phi_i^*({\bf x}_1)\phi_k^*({\bf x}_2)\frac{1}{r_{ij}}\phi_l({\bf x}_2) \phi_j({\bf x}_1)
    \end{equation}
\end{subequations}
where $\phi_i\left({\bf x}\right)$ represent one-electron spin-orbital basis and 
${\bf x}_i = (r_i, \sigma_i)$ represents electron spatial and spin coordinates.
Wave function anti-symmetry, originating from the Pauli exclusion principle, is enforced through the standard fermion anti-commutation relations $\{a_i,a_j^\dagger\}=\delta_{ij}$ and $\{a_i,a_j\}=\{a_i^\dagger,a_j^\dagger\}=0$. 

The second-quantized Hamiltonian represents indistinguishable electrons in molecules. However, quantum devices are composed of distinguishable spin systems. Mapping fermion field operators with raising and lowering Pauli operators will not generate a wave function satisfying the Pauli exclusion principle. Several methods have been developed to map fermion operators to spin systems~\cite{BKSF-2018, SeeleyBK2012, VerstraeteCirac, ParityMap}. Here, we will use the well-established Jordan-Wigner mapping method~\cite{JW-1928}. Jordan-Wigner mapping represents the electron orbital occupation directly in the qubit state $q_i = f_i\in\{0,1\}$, meaning the fermion basis directly maps to the qubit system. Anti-symmetry of the fermion operators results in a multiplicative phase factor upon the exchange of particles between orbitals that needs to be recovered in the mapping to spin operators. In Jordan Wigner mapping, raising $\sigma^+ = 1/2(\sigma_x + i \sigma_y)$ and lowing operators $\sigma^- = 1/2(\sigma_x -i \sigma_y)$ change the occupation, and the phase factor is recovered by strings of Pauli $\sigma_z$ operators. The fermion operators become
\begin{subequations}
\begin{equation}
    a_i^\dagger = \sigma_i^- \otimes \sigma_i^z\otimes ...\otimes \sigma_0^z
\end{equation}
\begin{equation}
    a_i^\dagger = \sigma_i^+ \otimes \sigma_i^z\otimes ...\otimes \sigma_0^z.
\end{equation}
\end{subequations}
Mapping the electronic Hamiltonian to a spin system using Jordan-Wigner enables the Hamiltonian to be simulated on the quantum computer. It is worth noting there are disadvantages to this mapping method; for instance, the non-locality of fermionic operators requires long strings of $\sigma_z$ operators when mapped to spins. This non-locality results in a larger number of entangling operations, which are the most error-prone operations on quantum hardware. 

For any fermion mapping method, the number of qubits scales with the number of orbitals, and coherent qubits are a limited resource on NISQ hardware. Previous quantum simulation demonstrations account for limited qubit resources by representing atomic orbitals in a minimal basis \cite{Kandala2017, OMalley2016, Colless2018, Lanyon2010, Hempel2018}. In electronic structure theory, a minimal basis can not properly capture the correlation effects, and utilizing a more complete basis set, like a correlation-consistent basis\cite{dunning-cc}, leads to more accurate energies. Working with large basis sets in the near-term will prove difficult since the resources are limited. By introducing methods to downfold select correlation effects into an effective Hamiltonian, it is possible to capture the correlation energy in a reduced space with variational quantum algorithms. 

\subsection{Variational Quantum Eigensolver}

A seminal demonstration of molecular Hamiltonian simulation on quantum hardware compared two quantum simulation algorithms: Iterative Quantum Phase Estimation (IQPE) and the Variational Quantum Eigensolver (VQE)\cite{OMalley2016}. The first method is a fully quantum routine that captures the energy eigenvalue in the phase of an auxiliary qubit after applying a unitary approximation of the Hamiltonian with algorithmic complexity bounds that suggest efficient solutions on a universal, fault-tolerant quantum computer.  NISQ devices are not fault-tolerant, have limited coherence times, and are prone to error. VQE is a heuristic, hybrid classical-quantum algorithm developed to compensate for these limitations and is more robust to error than the fully quantum IQPE algorithm. We implement the VQE algorithm using quantum circuit simulators to simulate molecular Hamiltonians in a downfolded active space.

Consider a wave function $\ket{\psi(\theta)}$ parameterized by a set $\{\theta_i\}$ of {\it independent} parameters. The variational principle states that 
\begin{equation}
    \matrixel{\psi(\theta)}{H}{\psi(\theta)} \geq E_g
\end{equation}
where $E_g$ is the ground-state energy of the Hamiltonian. The ground-state energy can be determined by optimizing the parameter set to find values that yield a minimum energy. The wave function ansatz can be generated on a quantum computer by applying a parameterized unitary operator
\begin{equation}
    \ket{\psi(\theta)} = U(\theta)\ket{\psi_0}
\end{equation}
to a simple trial wave function 
$\ket{\psi_0}$
that has good overlap to the genuine wave function. We assume that the Hartree-Fock wave function in the molecular-orbital basis is a good $\ket{\psi_0}$ for the chemically inspired ansatz~\cite{RomeroUCCSD}. After generating the wave function ansatz the expectation value of the Hamiltonian is needed to calculate the energy. A molecular Hamiltonian mapped to spin operators has the form 
\begin{equation}
    H = \sum_n h_n \prod_m \sigma_m
\end{equation}
and
\begin{equation}
    \matrixel{\psi(\theta)}{H}{\psi(\theta)} = \sum_n h_n \langle\psi(\theta)|\prod_m \sigma_m|\psi(\theta)\rangle
\end{equation}
so the energy can be directly calculated from a series of independent measurements measuring the expectation values for each term in the Hamiltonian. The cost of measuring  many terms of molecular Hamiltonians with large active spaces is a significant computational cost. In VQE, the optimization of the parameters happens on a classical computer. The quantum computer serves as an expectation value and total energy evaluation engine for the classical stochastic optimizer searching for the optimal set of parameters to generate the wave function ansatz. VQE simulations require many evaluations on the quantum device, due to the large number of parameters and the complexity of the parameter landscape.  

Critical to the success of VQE is an ansatz that captures the complexity of the wave-function and encapsulates the genuine ground state energy in the parameter space. Currently, there exist two classes of ansatz for VQE: hardware efficient and chemically inspired. Hardware efficient ansatz are composed of circuits with short depth with a limited number of gates, however, when used for molecular simulation have difficulties converging to a minimum~\cite{Kandala2017}. We therefore use a chemically-inspired UCCSD ansatz proposed in the original quantum simulation for chemistry research~\cite{peruzzo2014variational}. The UCCSD ansatz is defined using coupled-cluster theory and does not suffer from the convergence problems of the hardware efficient ansatz.  The coupled-cluster wave function 
\begin{equation}
    \ket{\psi_{CC}} = e^T\ket{\phi}
\end{equation}
is determined from the electron excitations from occupied to unoccupied orbitals within the active space. 
In the above equation the so-called reference function $\ket{\phi}$ is usually chosen as a Hartree-Fock determinant $\ket{\phi_{HF}}$.
If all excitations are included, the coupled-cluster wave function is equivalent to the exact configuration interaction wave function. Normally, the excitation operator $T$ is truncated at single and double excitations (CCSD)
\begin{subequations}
\begin{equation}
    T\simeq T_{SD} = T_1 + T_2,
\end{equation}
\begin{equation}
    T_1 = \sum_{ia} t_{ia} a_a^\dagger a_i,
\end{equation}
and
\begin{equation}
    T_2 = \sum_{ijab} t_{ijab} a_a^\dagger a_b^\dagger a_j a_i
\end{equation}
\end{subequations}
where $i,j,\ldots$ ($a,b,\ldots$) subscripts represent occupied (virtual) orbitals. Unfortunately, these operators are non-unitary. Non-unitary operators are unfit to perform on a quantum computer, generating wave functions that violate the variational principle. An alternative us the more computationally difficult unitary coupled-cluster (UCC) theory
\begin{equation}
    \ket{\psi_{UCC}} = e^{T-T^\dagger}\ket{\phi_{HF}}
\end{equation}
to parameterize the wave function. A first-order Trotter approximation of the time-evolution UCCSD operator~\cite{RevModPhys.92.015003} 
\begin{equation}
    U(t) = e^{-i\left(T-T^\dagger\right) t/\hbar}
\end{equation}
is used to construct the quantum circuit with $\hbar = t = 1$. Though the trotterized time-evolution UCCSD operator is not formally proven to be equivalent to the full UCCSD operator, in practice it is a sufficient operator to capture the correlation needed to converge to the ground-state energy variationally. The UCCSD time-evolved operator becomes computationally expensive for larger numbers of orbitals even when truncated to single and double excitations because the depth increases (rather significantly) with each additional excitation. Further improvements have recently been made on the UCCSD ansatz to capture correlation with shorter gate depth by carefully choosing the double excitations included in the ansatz~\cite{ADAPT_VQE, Lee2019}. Inspired by the use of second-order M\o ller-Plesset perturbation theory (MP2) coefficients as initial amplitudes in Ref.~\cite{RomeroUCCSD}, we further employ MP2 to get the most-important double excitations for a UCCS(D) operator, eliminating excitations with a MP2 amplitude below a certain threshold. To our knowledge, this is the first demonstration using MP2 for double excitation importance evaluations to construct an approximate UCCSD ansatz coupled with the amplitudes as initial optimization parameters. Running an algorithm for a larger active space on a quantum computer requires approximations in order to reduce the circuit complexity to within the coherence time of the device.

\section{Resource Reduction with DUCC}
\label{section:DUCC}
The DUCC formalism is predicated on the explicit decoupling excitations describing correlation effect in and outside of active space, i.e.,
\begin{equation}
\ket{\Psi_{DUCC}}=e^{\sigma_{\rm ext}} e^{\sigma_{\rm int}}\ket{\phi_{HF}} \;,
\label{ducc1}
\end{equation}
where $\sigma_{\rm int}$ and $\sigma_{\rm ext}$ are anti-Hermitian cluster operators defined as
\begin{eqnarray}
\sigma_{\rm int} &=& T_{\rm int}-T^{\dagger}_{\rm int}  \label{ints} \;, \\
\sigma_{\rm ext} &=& T_{\rm ext}-T^{\dagger}_{\rm ext} \label{exts} \;.
\end{eqnarray}
In the above formulas, $T_{\rm int}$ and $T_{\rm ext}$  CC-like cluster operators
producing excited configurations within and outside of the active space when acting on the reference function $\ket{\phi_{HF}}$, which in this case is the Hartree-Fock determinant.

The many-body analysis of the DUCC equations leads to the conclusion that for  $\sigma_{\rm int}$ and $\sigma_{\rm ext}$ satisfying DUCC equations:
\begin{equation}
 Qe^{-\sigma_{\rm int}}e^{-\sigma_{\rm ext}} H e^{\sigma_{\rm ext}}e^{\sigma_{\rm int}} \ket{\phi_{HF}} = 0   \;,  \label{ducceq}   
\end{equation}
where $Q$ represent the projection operator onto space orthogonal to the reference function $\ket{\phi_{HF}}$, 
the energy $E$, in contrast to standard expression
\begin{equation}
E=\bra{\phi_{HF}}e^{-\sigma_{\rm int}}e^{-\sigma_{\rm ext}} H e^{\sigma_{\rm ext}}e^{\sigma_{\rm int}} \ket{\phi_{HF}} \;,
\label{duccene}
\end{equation}
can be obtained by diagonalizing effective Hamiltonian
$\bar{H}_{\rm ext}^{\rm eff(DUCC)}$ in the corresponding active space (defined by projection operator $P+Q_{\rm int}$, where $P$ and $Q_{\rm int}$ are projection operators onto the reference function and  orthogonal determinants in the active space, respectively) 
\begin{equation}
        \bar{H}_{\rm ext}^{\rm eff(DUCC)} e^{\sigma_{\rm int}} \ket{\phi_{HF}} = E e^{\sigma_{\rm int}}\ket{\phi_{HF}}
\label{duccstep2}
\end{equation}
where
\begin{equation}
        \bar{H}_{\rm ext}^{\rm eff(DUCC)} = (P+Q_{\rm int}) \bar{H}_{\rm ext}^{\rm DUCC} (P+Q_{\rm int})
\label{equivducc}
\end{equation}
and
\begin{equation}
        \bar{H}_{\rm ext}^{\rm DUCC} =e^{-\sigma_{\rm ext}}H e^{\sigma_{\rm ext}} \;.
\label{duccexth}
\end{equation}
Formula (\ref{duccstep2}) forms a foundation for the DUCC downfolding formalism, where one constructs approximate many-body form of $\bar{H}_{\rm ext}^{\rm eff(DUCC)}$ using approximate form of 
$\sigma_{\rm ext}$ ($T_{\rm ext}$). Since, the $T_{\rm ext}$ operator contains higher-energy excitations, they can be approximated using various perturbative techniques. 

The process of approximating $\bar{H}_{\rm ext}^{\rm eff(DUCC)}$ involves three critical issues: (1) length of the commutator expansion for $e^{-\sigma_{\rm ext}}H e^{\sigma_{\rm ext}}$, (2) rank of many-body effects included in $\bar{H}_{\rm ext}^{\rm eff(DUCC)}$, 
and (3) approximate representation  of $T_{\rm ext}$. In this paper, we will follow the rudimentary approximation procedure explored in Ref.~\onlinecite{bauman2019downfolding},  where it was assumed that the 
second-order consistent form of $e^{-\sigma_{\rm ext}}H e^{\sigma_{\rm ext}}$ is used
\begin{equation}
e^{-\sigma_{\rm ext}}H e^{\sigma_{\rm ext}} \simeq H+[H_N,\sigma_{\rm ext}]+\frac{1}{2!}[[F_N,\sigma_{\rm ext}],\sigma_{\rm ext}] \;,
\label{happ1}
\end{equation}
where $F_N$ and $H_N$ represent Fock and Hamiltonian operators in normal product form. 
$T_{\rm ext}$ operator is expressed in terms of external part of the standard CCSD cluster operator, and one- and two-many-body terms are included in $\bar{H}_{\rm ext}^{\rm eff(DUCC)}$, i.e.,
\begin{equation}
\bar{H}_{\rm ext}^{\rm eff(DUCC)} \rightarrow \sum_{PQ} \chi^P_Q a_P^{\dagger} a_Q + \frac{1}{4} \sum_{P,Q,R,S} \chi^{PQ}_{RS} a_P^{\dagger} a_Q^{\dagger} a_S a_R \;,
\label{gammaph}
\end{equation}
where all summations run over active spin orbitals (designated here by $P,Q,\ldots$) and $\chi^{PQ}_{RS}$ represent anti-symmetrized  dressed two-electron integrals. This work utilizes the full form of
the one- and two-electron integrals, in contrast to Refs.~\onlinecite{bauman2019downfolding,bauman2019quantum}, which 
employed an orbital approximation of the spin-orbital $\chi$ tensor,  where it was assumed that Mulliken orbital-type  dressed integrals $({\bf P} {\bf Q}|\bf{R} \bf{S})$ are
obtained from  $\chi^{{\bf P}\alpha {\bf Q}\beta}_{{\bf R}\alpha {\bf S}\beta}$.

As mentioned earlier, the $\bar{H}_{\rm ext}^{\rm eff(DUCC)}$ depends only on the correlation effects related to the $\sigma_{\rm ext}$ operator and the fact that it operates only in the active space reduces the dimension of the problem compared to the full electronic Hamiltonian space involving whole spin-orbital space. As in Ref.~\onlinecite{bauman2020coupled}, we will study the downfolding procedures for  virtual orbitals only. In this way, active spaces employed in this paper will contain all occupied orbitals and small fractions of virtual orbitals. 

The dominating factors controlling the cost of calculating the downfolded Hamiltonian are 
the method used to provide information about the cluster operator and the size of the active 
space used to represent the downfolded Hamiltonian.  In this study, we use the CCSD method as 
a source of amplitudes, which scales as $n_o^2 n_v^4$ on the classical computer, where $n_o$ 
and $n_v$ refer to the total number of occupied and virtual orbitals. The classical CCSD 
calculation is the most computationally demanding step because if we follow the truncation 
outlined in Eq. \ref{happ1}, then many of tensor contractions defining the active-space 
representation of the downfolded Hamiltonian involve summations over a small set of virtual 
active orbitals. The associated cost of forming the downfolded Hamiltonian is significantly 
smaller than $n_o^2 n_v^4$ and scales roughly as $n_o^4 n_v^2$, assuming that the number of 
active virtual orbitals is significantly smaller than the total number of occupied orbitals.

In the following part of this paper, we will invoke the VQE algorithms to minimize the functional
\begin{equation}
    \matrixel{\psi(\theta)}{\bar{H}_{\rm ext}^{\rm eff(DUCC)}}{\psi(\theta)} 
\end{equation}
in the active space. This approach allows one to significantly reduce the number of parameters involded in the VQE minimization procedure and focus on these which are critical to describe static correlation effects stemming from a given active space. 
In this situation the set of  $\{\theta_i\}$  {\it independent} parameters corresponds to the subset of  excitations within the active space.

\section{DUCC Benchmarking Using Quantum Circuit Simulations}

\begin{table*}[bt]
  \caption{Resource requirements for simulation: number of qubits, number of single and double excitations for molecule in the active space, and gate depth for UCCSD circuit. MP2 excitations calculated with a cutoff threshold of $10^{-5}$.}
  \label{Table:CircuitParam}
  \begin{tabular}{|l|c|ccccc|}
    \hline
     \bf{Molecule} &\bf Number of Orbitals & \bf{$N_q$} & \bf{All Excitations} & \bf{Depth UCCSD} & \thead{MP2 Excitations} & \bf{Depth UCCS(D)} \\
    \hline
    $\text{H}_2$ & 4 & 8 & 15 & 1,021 & 11 & 605\\
     & 5 & 10 & 24 & 2,049 & 16 & 1,089\\
     & 6 & 12 & 35 & 3,581 & 19 & 1,341\\
     & 7 & 14 & 48 & 5,713 & 22 & 1,633\\
     & 8 & 16 & 63 & 8,541 & 25 & 1,965\\
     & 9 & 18 & 80 & 12,161 & 28 & 2,337\\
     & 10 & 20 & 99 & 16,669 & 35 & 3,613\\
    \hline
    $\text{Li}_2$ & 7 & 14 & 204 & 25,537 & 56 & 5,153\\
    & 8 & 16 & 315 & 44,061 & 89 & 10,125\\
    & 9 & 18 & 450 & 69,361 & 124 & 15,873\\
    & 10 & 20 & 609 & 102,397 & 141 & 19,229\\
    & 11 & 22 & 792 & 144,129 & 206 & 32,305\\
    & 12 & 24 & 999 & 195,517 & 265 & 45,965\\
    & 13 & 26 & 1,230 & 257,521 & 324 & 60,577\\
    & 14 & 28 & 1,485 & 331,101 & 383 & 76,845\\
    \hline
  \end{tabular}
\end{table*}

The ability to capture dynamical correlation effects in a downfolded active space with DUCC is evaluated using quantum circuit simulation for $\text{H}_2$ and $\text{Li}_2$ using the cc-pVTZ basis\cite{Prascher2011}. Qiskit\cite{Qiskit} quantum simulation software is used to encode the molecular orbitals onto the spin operators using the Jordan-Wigner transform and construct a circuit that, in principle, can be executed on a quantum processing unit (QPU). For our simulations we used the state-vector simulator available in Qiskit, as the circuit depth and qubit count exceeds what is available experimentally. Essentially, quantum circuit simulators execute the algorithm and are a tool for determining sources of algorithmic error such as device noise, shot noise or error from trotter decomposition on the variational ansatz.  Numerous quantum circuit simulators becoming available as open-source computational libraries, and we found Qiskit to be an accessible and complete software library to simulate the VQE algorithm for molecular Hamiltonians.

\begin{figure}[t]
	\includegraphics[width = 2in]{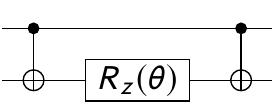}
	\caption{Circuit for operator $U = \text{exp}(-i\theta\sigma_z^1\sigma_z^2/2)$}
	\label{fig:circuit}
\end{figure}
When constructing a quantum algorithm for molecular Hamiltonians, there are two resources to consider: the number of qubits and the number of quantum operations or gates that need to be performed. Using standard fermionic to spin mapping techniques, the number of qubits $N_q$ is equal to the number of spin orbitals $M$, and quantum circuit simulation scales $2^M$. Working in a reduced active space where the external excitations are downfolded into an effective Hamiltonian reduces the number of qubits needed to encode the molecular Hamiltonian. An additional, considerable component of quantum algorithm simulation is the number of operations (gates) applied to the quantum devices. These operations are classified into single-qubit gates and multi-qubit (entangling) gates. Circuit synthesis for spin operators in the Hamiltonian, exponentiated in the unitary time-evolution operator, consists of single-qubit rotations and an entangling operation known as a CNOT~\cite{NandC}. For example a two-spin Hamiltonian $H = J\sigma_z^1\sigma_z^2$ that can be encoded as a unitary operation for a quantum computer using the time-evolution operator $U = \text{exp}(-i\theta/2\sigma_z^1\sigma_z^2)= \text{exp}(-iJt\sigma_z^1\sigma_z^2/\hbar)$ where $\hbar = t = 1$ for a simplistic demonstration. The unitary is encoded by entangling two qubits with a CNOT, performing a z-rotation, and applying an additional CNOT shown in Fig~\ref{fig:circuit}. Hamiltonians with $\sigma_x$ and $\sigma_y$ operators are treated in the same way, after applying a single-qubit rotation into the x or y basis accordingly. The circuit structure is important when working with the UCCSD ansatz since the operator takes the form of a spin Hamiltonian after the qubit mapping is applied. Each additional excitation in the ansatz comes with a significant overhead of gates that need to be applied to the qubits. For large active spaces, the computational cost of running our circuits on a QPU is high due to the large circuit volume $V = N_q *D$, where D is the circuit depth. For this reason, it is worth understanding the computational cost, from a practical perspective, for quantum circuit simulation of increasingly large active spaces. The resource requirements for the quantum circuit and optimization problem are included in Table~\ref{Table:CircuitParam}. Consider the difference between H$_2$ with 4 orbitals and H$_2$ with 10 orbitals, each additional virtual orbital requires 2 qubits and the number of excitations grows polynomially with added virtual orbitals. Additional excitations require more layers of gates.  Additional virtual orbitals have a significant effect on the circuit volume as well as the dimensionality of the optimization problem. As such, downfolding higher virtual orbitals using DUCC coupled with our perturbative doubles approach provides a significant computational advantage.
\begin{figure}[t]
	\includegraphics[trim= 1.1in 1.45in 1in 0.5in, clip,  width = 0.45\textwidth]{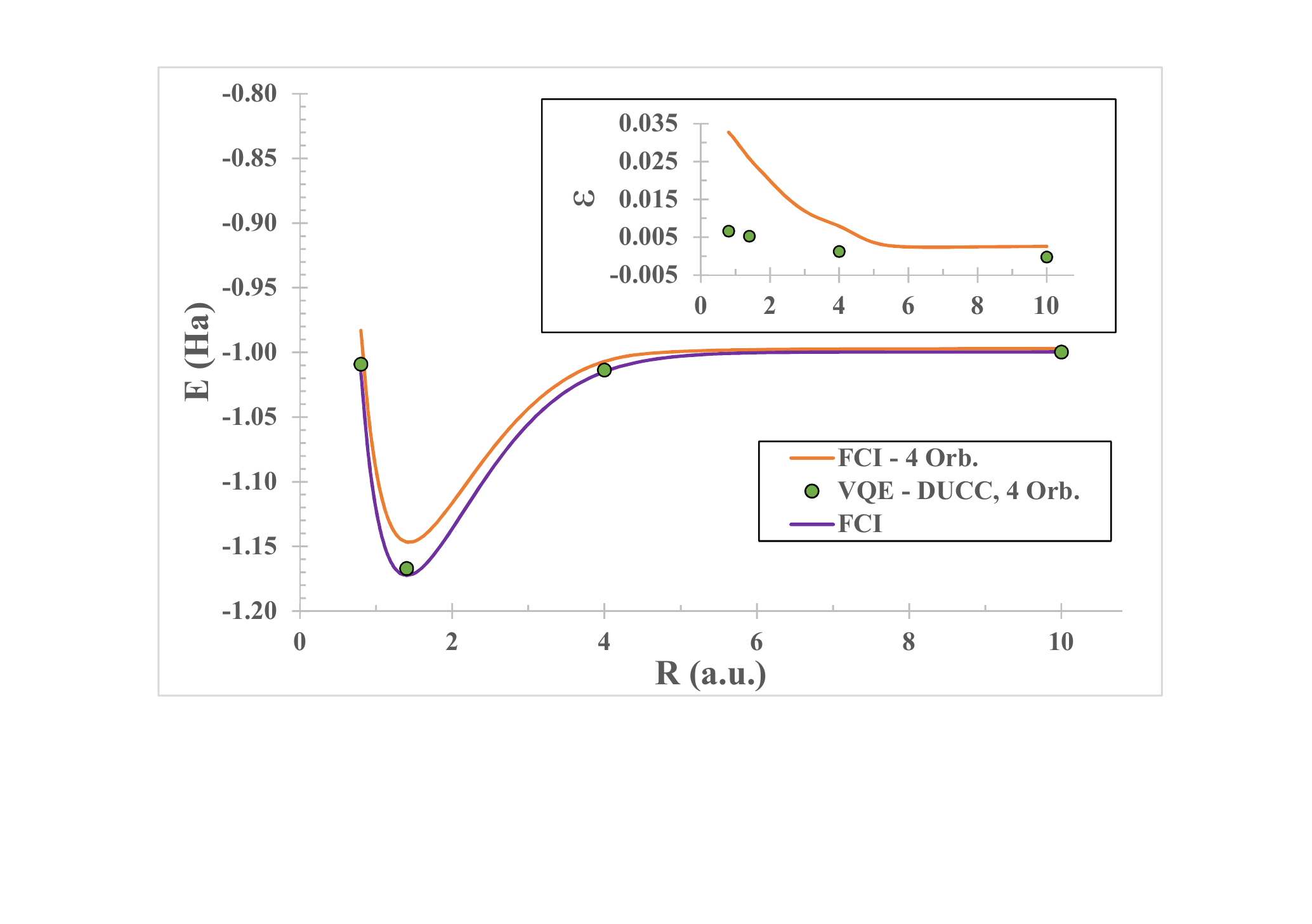}
	\caption{Ground state energy of $\text{H}_2$ with the cc-pvtz basis. The errors (inset), $\epsilon$, of the VQE-DUCC and the active-space FCI calculations are with respect to the all-orbital FCI energies.}
	\label{fig:h2}
\end{figure}

A starting point to benchmark the VQE-DUCC method is with a popular molecule for near-term quantum computers $\text{H}_2$. Energy evaluation of $\text{H}_2$, in a minimum basis, has been conducted on QPUs with variational algorithms\cite{Hempel2018,Lanyon2010, OMalley2016, Colless2018}. We employ the cc-pVTZ basis set with Cartesian angular functions and consider an active space of only four orbitals when constructing the DUCC Hamiltonian, a significant reduction compared to the 30 orbitals required for the full calculation with the bare Hamiltonian. The one and two-body integrals of the effective DUCC Hamiltonian are provided in the Appendix. We compute the energy of the DUCC Hamiltonian using VQE. The UCCSD ansatz for VQE includes all single and double excitations within the active space, and the circuit is simulated using a state-vector simulator on a classical computer. We employ the non-gradient based COBYLA~\cite{cobyla} optimizer for constrained optimization problems as the {\it classical} outer loop for the hybrid-variational quantum algorithm. The energy converged after $\mathcal{O}(10^2)$ calls to the optimizer and agreed with the exact diagonalization of the effective Hamiltonian to a tiny fraction of a millihartree. Energies from the active-space FCI calculation with the bare Hamiltonian and VQE with the DUCC Hamiltonian are compared to the all-orbital FCI results in Figure~\ref{fig:h2}. For the $\text{H}_2$ case, the VQE-DUCC approach is able to recover most of the correlation energy downfolding the higher virtuals into the 4 orbitals used in the active space, accurately recovering the FCI potential energy surface.
    
\begin{figure}[t]
	\includegraphics[trim= 0.9in 5.3in 1.4in 0.95in, clip,  width = 0.45\textwidth]{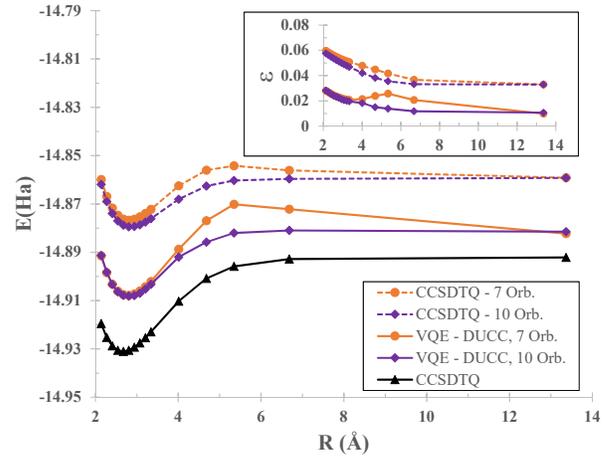}
	\caption{Ground state energy of $\text{Li}_2$ with the cc-pvtz basis. The errors (inset), $\epsilon$, of the VQE-DUCC and the active-space CCSDTQ calculations are with respect to the 60-orbital CCSDTQ energies.}
	\label{fig:li2}
\end{figure}

\begin{figure}[t]
	\includegraphics[trim= 0.4in 0.9in 0.4in 0.9in, clip,  width = 0.45\textwidth]{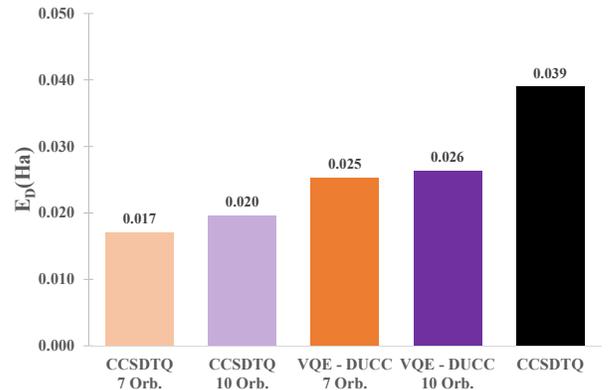}
	\caption{Dissociation energies E$_D$ of Li$_2$.}
	\label{fig:li2_de}
\end{figure}

Lithium is a crucial element at the center of current battery research. We investigate the diatomic lithium molecule $\text{Li}_2$ to test DUCC and VQE in larger active spaces. For the $\text{Li}_2$ system, we considered two active spaces consisting of seven and ten orbitals to construct the DUCC Hamiltonian while utilizing the cc-pVTZ basis set with spherical-harmonic angular functions. For comparison, 60-orbital and analogous active-space CCSDTQ calculations were run with the bare Hamiltonian since CCSDTQ is effectively exact for $\text{Li}_2$. Since the DUCC is not exact in practice, we have to be mindful of the choice of the active space. Too small of an active space can result in nonphysical barriers, as seen in Figure~\ref{fig:li2}. An active space of 10 orbitals is found to be sufficient to recover enough of the correlation such that these barriers vanish. In Figure~\ref{fig:li2_de} we show the dissociation energies. While downfolding improves the dissociation energy, it is not able to recover the full 60-orbital CCSDTQ value. The dissociation energy seems to be less sensitive to the number of orbitals utilized in the active space. We will discuss the reasons for this later in this section.
 
Constructing the circuit in these active spaces starts to become a challenging computational task as the number of included excitations, and thus optimization parameters begin to grow Table~\ref{Table:CircuitParam}. First, simulating a circuits with a large volume is computationally difficult and time-consuming. Second, a larger number of optimization parameters requires substantially more iterations to converge to minimum energy, and the computationally expensive circuit needs to be simulated many times to converge to a minimum energy within chemical accuracy for a {\it well-behaved} optimization surface. The number of iterations or runs needed for an ansatz with a ridged or barren optimization landscape is even greater. Thus, even the variational quantum algorithm succumbs to the balance between accuracy and complexity. 
    
To make the VQE simulation more tractable in larger active spaces, we make an approximation to the UCCSD ansatz. We compute the MP2 coefficients for the double excitations within the active space. Double excitations with MP2 amplitudes below a threshold of $10^{-5}$ are discarded, and we construct the UCCS(D) ansatz with the remaining perturbative excitations. This approximation decreases the number of optimization parameters and the gate depth significantly, therefore, the computational time for each circuit simulation is reduced by orders of magnitude Table~\ref{Table:CircuitParam}. The UCCS(D) ansatz yields energies within chemical accuracy when compared to the values obtained from the exact diagonalization of the Hamiltonian. Using the UCCS(D) ansatz is also beneficial for running on QPU since it can significantly decrease the number of gates needed to construct the ansatz. The optimizer made $\mathcal{O}(10^3)$ calls to converge for both the 7 orbital case and 10 orbital diatomic lithium with the MP2 amplitudes as an initial guess. Using these amplitudes as a preliminary guess for the optimizer reduces the number of calls to the quantum computer by orders of magnitude~\cite{RomeroUCCSD}.  In the case where there was no initial guess, the optimizer required $\mathcal{O}(10^4)$ calls to converge, which substantiates the importance of the initial guess for hybrid quantum algorithms. Using perturbation theory like MP2 for the ansatz and initial guess reduces the number of gates needed to construct the variational quantum circuit, leads to better optimizer performance with reduced hyper-surface dimensionality, and reduces the number of evaluations needed for to find the minimum energy.

How well the DUCC method performs relies on how well the
approximations made in defining the downfolded Hamiltonian describe the dynamical correlation effects. It is
important to emphasize that if the exact form of the downfolded 
Hamiltonian is utilized, then the exact energy would be obtained by diagonalizing the downfolded Hamiltonian in the active space, 
independent of the size of the active space. Although the exact form 
is unobtainable, the approximations laid out in Section \ref{section:DUCC} 
provide a reliable procedure for incorporating, mainly dynamical, correlation
effects into a dimensionality reduced effective Hamiltonian. It is apparent 
the DUCC Hamiltonian improves the agreement with FCI energies at each point 
along a potential energy curve compared to the bare Hamiltonian in the same 
active space, but best performs at stretch bond lengths 
(strongly correlated regime). This is rationalized by considering perturbative 
energy denominators, 
\begin{equation}
\frac{1}{\epsilon_i+\epsilon_j+\ldots - \epsilon_b - \epsilon_a}\;,
\label{eneden}
\end{equation}
where $\epsilon$'s are Hartree-Fock orbital energies. When these denominators 
become relatively large, as in the case of external excitations in instances of 
strong correlation, then perturbative expansions or low-rank CC methods (CCSD) 
should be sufficient to provide an accurate description of these excitations.

\begin{figure}[t]
	\includegraphics[trim= 0.95in 4.6in 1.42in 1.6in, clip,  width = 0.45\textwidth]{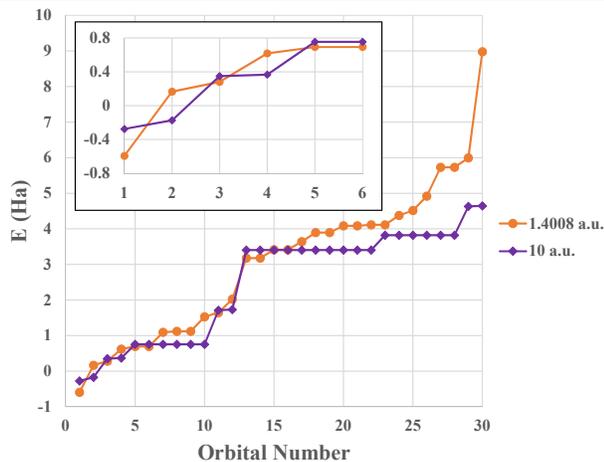}
	\caption{Hartree-Fock orbital energies corresponding to the equilibrium (1.4008 a.u.) and 
	stretch (10 a.u.) bond lengths for H$_{2}$. }
	\label{fig:H2orbene}
\end{figure}

In the case of H$_{2}$, stretched bond lengths are characterized by larger 
energy gaps between the highest energy orbital in the active space and 
the lowest-energy orbital outside of the active space, as seen in Figure
\ref{fig:H2orbene}. In addition, the 
overall range of orbital energies in the active space is smaller than in 
the case of the equilibrium bond length. Therefore, the internal and 
external excitations are well disjointed, and the active space ought to be 
more appropriate for the stretch bond length, which is indeed the case as
confirmed by examining the leading cluster amplitudes in 
Table \ref{H2excitations}. At 10 a.u., the active space encapsulates the 
dominant excitations, and the external excitation manifold is relatively 
small and well described by the DUCC Hamiltonian, reproducing the FCI 
energy to a fraction of a millihartree, an order of magnitude smaller 
than the bare Hamiltonian is the same active space. The DUCC Hamiltonian 
at equilibrium provides significant improvement to the bare Hamiltonian 
in the same active space, reducing the error compared to FCI by a factor 
of about 5. This is accomplished despite the internal and external 
excitations not being well disjointed, as seen in Figure \ref{fig:H2orbene} 
and reinforced by the leading cluster amplitudes in Table \ref{H2excitations},
which is why the error is larger near the equilibrium than the stretch bond
lengths. Longer commutator expansions and the consideration of higher-order
components of the DUCC Hamiltonian will further reduce this error. 

\begin{center}
\begin{table}[]
    \centering
\caption{The five largest cluster amplitudes for the equilibrium (1.4008 a.u.)
    and stretch (10 a.u.) bond lengths of H$_{2}$ from the corresponding CCSD calculations.}
\begin{tabular}{ l c }
Excitation Character & |Amplitude| \\
\hline \\ [-3mm]
\multicolumn{2}{c}{1.4008 a.u.} \\ [1mm]
1$\alpha$ 1$\beta$ $\rightarrow$ 4$\alpha$ 4$\beta$   &  0.055 \\
1$\alpha$ 1$\beta$ $\rightarrow$ 2$\alpha$ 4$\beta$ / 4$\alpha$ 2$\beta$  &  0.044 \\
1$\alpha$ 1$\beta$ $\rightarrow$ 5$\alpha$ 5$\beta$ / 6$\alpha$ 6$\beta$  &  0.042 \\
1$\alpha$ 1$\beta$ $\rightarrow$ 2$\alpha$ 2$\beta$   &  0.039 \\
1$\alpha$ 1$\beta$ $\rightarrow$ 3$\alpha$ 3$\beta$   &  0.035 \\[1mm]
\multicolumn{2}{c}{10 a.u.} \\[1mm]
1$\alpha$ 1$\beta$ $\rightarrow$  2$\alpha$ 2$\beta$  &   0.997 \\
1$\alpha$ / 1$\beta$ $\rightarrow$ 3$\alpha$ / 3$\beta$   &   0.152 \\
1$\alpha$ 1$\beta$ $\rightarrow$  2$\alpha$ 4$\beta$ / 4$\alpha$ 2$\beta$  &   0.146 \\
1$\alpha$ / 1$\beta$ $\rightarrow$ 11$\alpha$ / 11$\beta$   &   0.027 \\
1$\alpha$ 1$\beta$ $\rightarrow$  2$\alpha$ 12$\beta$ / 12$\alpha$ 2$\beta$  &   0.026
\end{tabular}
\label{H2excitations}
\end{table}
\end{center}

\begin{figure}[t]
	\includegraphics[trim= 0.95in 4.9in 1.48in 1.5in, clip,  width = 0.45\textwidth]{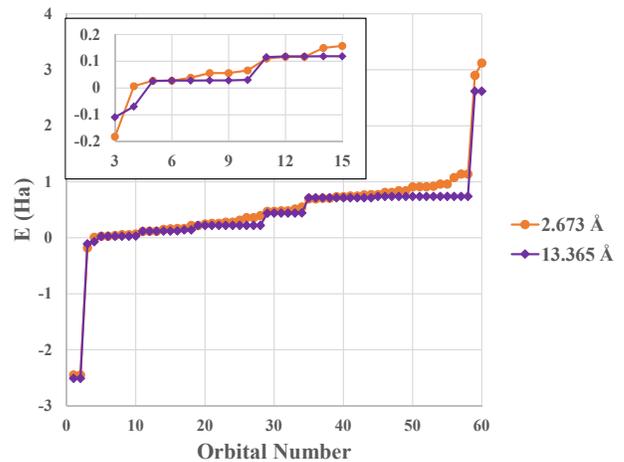}
	\caption{Hartree-Fock orbital energies corresponding to the equilibrium (2.673 \AA) and 
	stretch (13.365 \AA) bond lengths for Li$_{2}$ from the corresponding CCSD calculations. }
	\label{fig:Li2orbene}
\end{figure}

For Li$_{2}$, the DUCC Hamiltonian provides a significant improvement at 
each point along the potential energy curve over the CCSDTQ calculation in the 7 and 10 orbital representations, and DUCC captures enough correlation in the reduced active space needed to improve the dissociation energy despite the somewhat undesirable
conditions for utilizing the downfolding procedure, as shown in Figures~\ref{fig:li2} and \ref{fig:li2_de}. As seen in Figure
\ref{fig:Li2orbene}, the overall orbital energy range varies less than 
the case of H$_{2}$ as almost the entire set of virtual orbitals is below 
1 Hartree in energy. Consequently, it is hard to define an active space where there 
is a good division between internal and external excitations as seen in 
Table \ref{Li2excitations}. Even excitations from the HOMO to the 
highest-energy unoccupied orbital are nonnegligible.  Still, the 
DUCC Hamiltonian provides improvement over the bare Hamiltonian in the 
same active space with a reduction of error in total energies by a 
factor of 2 at the shorter bond lengths and up to a factor of 3.5 at the longer bond lengths, as well as a reduction of error in the dissociation energy 
from 19--22 milliHartree to 12--14 milliHartree for the two active spaces (see Fig. \ref{fig:li2_de}). 
The DUCC Hamiltonian 
performs best at the stretched bond lengths for the same rationalization 
as with H$_{2}$. The inadequacy of the active spaces to capture important correlation effects at shorter bond lengths is also reflected in the equilibrium bond lengths of the different methods. The DUCC Hamiltonians provide a better agreement of the equilibrium bond length with the full CCSDTQ value than the active-space CCSDTQ calculations with the bare Hamiltonian in their respective active spaces, however they are still 0.1--0.15 \mbox{\normalfont\AA}
longer than the full CCSDTQ results. A longer commutator expansion that leads to more 
accurate DUCC Hamiltonians will play a significant role in further 
reducing the errors, which will be explored in future studies.

\begin{center}
\begin{table}[]
    \centering
\caption{The five largest cluster amplitudes for the equilibrium (2.673 \AA) and 
	stretch (13.365 \AA) bond lengths of Li$_{2}$.}
\begin{tabular}{ l c }
Excitation Character & |Amplitude| \\
\hline \\ [-3mm]
\multicolumn{2}{c}{2.673 \AA} \\ [1mm]
3$\alpha$ 3$\beta$ $\rightarrow$ 5$\alpha$ 5$\beta$ / 6$\alpha$ 6$\beta$ &  0.095 \\
3$\alpha$ 3$\beta$ $\rightarrow$ 5$\alpha$ 13$\beta$ / 6$\alpha$ 12$\beta$ /
13$\alpha$ 5$\beta$ / 12$\alpha$ 6$\beta$ & 0.081 \\
3$\alpha$ 3$\beta$ $\rightarrow$ 12$\alpha$ 12$\beta$ / 13$\alpha$ 13$\beta$ &  0.074 \\
3$\alpha$ 3$\beta$ $\rightarrow$ 11$\alpha$ 11$\beta$ &  0.072 \\
3$\alpha$ 3$\beta$ $\rightarrow$ 4$\alpha$ 4$\beta$ &  0.069 \\[1mm]
\multicolumn{2}{c}{13.365 \AA} \\[1mm]
3$\alpha$ 3$\beta$ $\rightarrow$ 4$\alpha$ 4$\beta$ &   0.984 \\
3$\alpha$ / 3$\beta$ $\rightarrow$ 17$\alpha$ / 17$\beta$ &   0.137 \\
3$\alpha$ 3$\beta$ $\rightarrow$ 4$\alpha$ 18$\beta$ / 18$\alpha$ 4$\beta$ &   0.133 \\
3$\alpha$ / 3$\beta$ $\rightarrow$ 5$\alpha$ / 5$\beta$ &   0.028 \\
3$\alpha$ 3$\beta$ $\rightarrow$ 4$\alpha$ 10$\beta$ / 10$\alpha$ 4$\beta$ &   0.019 
\end{tabular}
\end{table}
\label{Li2excitations}
\end{center}


\section{Conclusion}
We have demonstrated the efficacy of using active-space DUCC downfolded Hamiltonians given by second-order consistent commutator expansion (Eq.(\ref{happ1}) with the Variational Quantum Eigensolver (VQE) to model the potential energy surfaces of molecules on a quantum computer using the full molecular orbital basis. In all calculations, only one and two many-body components of DUCC downfolded Hamiltonians were considered. We used the VQE algorithms to minimize DUCC Hamiltonians using unitary CC expansion with singles and doubles acting in the  active space. 
This procedure was applied to two types of systems: (1) the active space is well defined and external amplitudes can be obtained with high accuracy (the H$_2$ system) and (2) the ideal active space is too large and only its sub-spaces can be considered in practical calculations (the Li$_2$ molecule). 

In the case of  the H$_2$ molecule, we were able to obtain a very good agreement of VQE-DUCC energies with FCI ones for all internuclear geometries of H$_2$ considered here. For the  Li$_2$ case, the HF molecular basis does not lend
itself to easily define an active space as practically all virtual orbitals should be deemed active. Despite this obstacle, the DUCC Hamiltonian provides sufficient improvement (especially for the active space consisting of ten active orbitals) over the bare Hamiltonian in the same size active spaces when compared to the all-orbital benchmark
CCSDTQ calculations. The DUCC Hamiltonian best performs at large internuclear Li-Li separations where the active spaces encapsulate the most important correlation effects as supported by examining the leading excitations. The 
discrepencies between the VQE-DUCC and the benchmark CCSDTQ calculations can be largely attributed to the truncation of the commutator expansion used to define the DUCC Hamiltonian. Another important factor contributing to  the performance of the UCCSD  formalism is the choice of the orbitals. 


These observations give us a strong motivation towards further improving the approximations used to define the DUCC Hamiltonian, including incorporating higher-order terms and excitations in the commutator expansions. The utilization of natural orbitals will additionally provide a natural way of separating static and dynamical correlation effects and more reliable definitions of active spaces.

\section{Acknowledgement}
This  work  was  supported  by  the "Embedding Quantum Computing into Many-body Frameworks for Strongly Correlated  Molecular and Materials Systems" project, 
which is funded by the U.S. Department of Energy(DOE), Office of Science, Office of Basic Energy Sciences, the Division of Chemical Sciences, Geosciences, and Biosciences.
Calculations have been performed using computational resources  at the Pacific Northwest National Laboratory (PNNL).
PNNL is operated for the U.S. Department of Energy by the Battelle Memorial Institute under Contract DE-AC06-76RLO-1830. This research used resources of the National Energy Research Scientific Computing Center (NERSC). NERSC is a U.S. Department of Energy Office of Science User Facility operated at LBNL under Contract No. DE-AC02-05CH11231. 

\bibliographystyle{apsrev4-1}
\bibliography{references}

\appendix
\begin{table*}[bt]
  \caption{DUCC Integrals for H$_2$.\footnote{Only unique integrals are listed. The full set of integrals can be 
  obtained by exploiting the symmetry of the integrals: (i|j) = (j|i) and (ij|kl) = (ji|lk) = (kl|ij) = (lk|ji).}}
  \label{Table:H2 Int}
  \begin{tabular}{|cccc|rrrr|}
    \hline
     i  & j & k & l & \multicolumn{1}{c}{0.8 a.u.} & \multicolumn{1}{c}{1.4008 a.u.} & \multicolumn{1}{c}{4 a.u.} & \multicolumn{1}{c}{10 a.u.} \\
     \hline
1 &	1 &	- &	- &	-1.5133089437 &	-1.2528398693 &	-0.7774913856 &	-0.5884671771 \\
2 &	2 &	- &	- &	-0.4038274345 &	-0.4695131026 &	-0.6613171035 &	-0.5879007924 \\
3 &	1 &	- &	- &	-0.1435962544 &	-0.1311715782 &	-0.0825865887 &	-0.0754739366 \\
3 &	3 &	- &	- &	-0.3974760145 &	-0.3575220896 &	-0.1967048543 &	-0.1005238389 \\
4 &	2 &	- &	- &	-0.1953866501 &	-0.2102172352 &	-0.1254423710 &	-0.0744086544 \\
4 &	4 &	- &	- &	-0.1764021745 &	-0.2484788837 &	-0.1733529055 &	-0.0830028510 \\
1 &	1 &	1 &	1 &	 0.7724268885 &	 0.6378951380 &	 0.3931534153 &	 0.3149555522 \\
1 &	1 &	1 &	3 &	 0.1389552980 &	 0.1249166955 &	 0.0792039077 &	 0.0736003818 \\
1 &	2 &	1 &	2 &	 0.0173294476 &	 0.0379391632 &	 0.1510484416 &	 0.2150501447 \\
1 &	2 &	1 &	4 &	 0.0312395566 &	 0.0538042833 &	 0.0669877909 &	 0.0741307752 \\
1 &	3 &	1 &	3 &	 0.0509714461 &	 0.0525096611 &	 0.0427308654 &	 0.0451139499 \\
1 &	4 &	1 &	4 &	 0.0672304776 &	 0.0916251037 &	 0.0456196352 &	 0.0452088106 \\
1 &	1 &	2 &	2 &	 0.3084029922 &	 0.3375192457 &	 0.3787534458 &	 0.3154484987 \\
1 &	1 &	2 &	4 &	 0.1140154595 &	 0.1333117072 &	 0.0959646324 &	 0.0750813643 \\
1 &	2 &	2 &	1 &	 0.0179792300 &	 0.0404576306 &	 0.1543354457 &	 0.2152992614 \\
1 &	2 &	2 &	3 &	-0.0196992413 &	 0.0216228218 &	 0.0525936784 &	 0.0763401704 \\
1 &	3 &	2 &	2 &	 0.0105520419 &	 0.0225835506 &	 0.0758172817 &	 0.0770005829 \\
1 &	3 &	2 &	4 &	 0.0185507905 &	 0.0341108797 &	 0.0480893106 &	 0.0466232945 \\
1 &	4 &	2 &	1 &	 0.0321407532 &	 0.0559186698 &	 0.0677152847 &	 0.0742649617 \\
1 &	4 &	2 &	3 &	-0.0210203222 &	 0.0098161740 &	 0.0412122702 &	 0.0465235061 \\
1 &	1 &	3 &	3 &	 0.3488131162 &	 0.3440317944 &	 0.3044062264 &	 0.2480816635 \\
1 &	2 &	3 &	2 &	-0.0193269488 &	 0.0209721806 &	 0.0534761516 &	 0.0763652312 \\
1 &	2 &	3 &	4 &	 0.0043553793 &	 0.0251429464 &	 0.0983731536 &	 0.1483575642 \\
1 &	3 &	3 &	1 &	 0.0444155483 &	 0.0489499190 &	 0.0453031421 &	 0.0474929818 \\
1 &	3 &	3 &	3 &	 0.0229055027 &	 0.0301200420 &	 0.0395054275 &	 0.0416682630 \\
1 &	4 &	3 &	2 &	-0.0198678734 &	 0.0084614128 &	 0.0419772903 &	 0.0474484327 \\
1 &	4 &	3 &	4 &	 0.0081514280 &	 0.0343892750 &	 0.0350286932 &	 0.0409012271 \\
1 &	1 &	4 &	4 &	 0.4693710195 &	 0.4796377358 &	 0.3012543120 &	 0.2483925640 \\
1 &	2 &	4 &	3 &	 0.0034195825 &	 0.0239347316 &	 0.0979099327 &	 0.1482143806 \\
1 &	3 &	4 &	2 &	 0.0187674424 &	 0.0342574260 &	 0.0483911099 &	 0.0474050376 \\
1 &	3 &	4 &	4 &	 0.0451667198 &	 0.0695945403 &	 0.0453344341 &	 0.0420494471 \\
1 &	4 &	4 &	1 &	 0.0662576180 &	 0.0928641097 &	 0.0479282470 &	 0.0468829748 \\
1 &	4 &	4 &	3 &	 0.0057348752 &	 0.0321351355 &	 0.0347261691 &	 0.0407449880 \\
2 &	2 &	2 &	2 &	 0.2741917775 &	 0.2903110488 &	 0.3703764515 &	 0.3145786150 \\
2 &	2 &	2 &	4 &	 0.0291801314 &	 0.0434143395 &	 0.0846518034 &	 0.0733936961 \\
2 &	3 &	2 &	3 &	 0.0491679076 &	 0.0393026982 &	 0.0360618956 &	 0.0459572261 \\
2 &	4 &	2 &	4 &	 0.0369961165 &	 0.0523031625 &	 0.0549684077 &	 0.0453123001 \\
2 &	2 &	3 &	3 &	 0.2536160173 &	 0.2614638476 &	 0.2901670066 &	 0.2473719020 \\
2 &	3 &	3 &	4 &	-0.0075230699 &	 0.0181076777 &	 0.0244436264 &	 0.0418925938 \\
2 &	4 &	3 &	3 &	 0.0354488082 &	 0.0394947099 &	 0.0492995405 &	 0.0406881984 \\
2 &	2 &	4 &	4 &	 0.2894064380 &	 0.3146952951 &	 0.2945494072 &	 0.2475110205 \\
2 &	3 &	4 &	3 &	-0.0075230699 &	 0.0181076777 &	 0.0244436264 &	 0.0418925938 \\
2 &	4 &	4 &	4 &	 0.0725280665 &	 0.0942192106 &	 0.0488131745 &	 0.0408456219 \\
3 &	3 &	3 &	3 &	 0.2671135355 &	 0.2661184719 &	 0.2585481275 &	 0.2125686827 \\
3 &	4 &	3 &	4 &	 0.0076848386 &	 0.0216780341 &	 0.0724820536 &	 0.1123997215 \\
3 &	3 &	4 &	4 &	 0.2863619808 &	 0.2923360211 &	 0.2502475280 &	 0.2123756056 \\
4 &	4 &	4 &	4 &	 0.3740766949 &	 0.4049606919 &	 0.2549328171 &	 0.2123359658 \\
        \hline
  \end{tabular}
\end{table*}

\begin{table*}[bt]
  \caption{Optimal Coefficients for H$_2$}
  \label{Table:H2 Excit}
  \begin{tabular}{|c|rrrr|}
    \hline
     \multicolumn{1}{c}{Excitation} & \multicolumn{1}{c}{0.8 a.u.} & \multicolumn{1}{c}{1.4008 a.u.} & \multicolumn{1}{c}{4 a.u.} & \multicolumn{1}{c}{10 a.u.} \\
     \hline
     $t_{1\alpha 2\alpha}$ & -0.00169188 & 0.00021634 & 0.0117512 &-0.05214657\\
     $t_{1\alpha 3\alpha}$ & 0.00123214& 0.00672613 & 0.0704723 & 0.10784842\\
     $t_{1\alpha 4\alpha}$ & -0.00073229 & 0.00010555 & 0.00216901 & -0.00748003\\
     $t_{1\beta 2\beta}$ & 0.00065677 & 0.00004992 & -0.01133357 & 0.05239833\\
     $t_{1\beta 3\beta}$ & 0.00087783 & 0.00694961 & 0.06994873 &   0.01074307\\
     $t_{1\beta 4\beta}$ & 0.00074183 & -0.00039769 & -0.00241246 & 0.00747299\\
     $t_{1\alpha 1\beta 2\alpha 2\beta}$ & -0.01179497 &  -0.039163 & -0.48799278 & -0.77241219\\
     $t_{1\alpha 1\beta 2\alpha 3\beta}$ & -0.00039156 & 0.00015069& 0.000216463 &  -0.00283207\\
     $t_{1\alpha 1\beta 2\alpha 4\beta}$ & -0.01812841 &  -0.0438853&  -0.12975184 &  -0.14512297\\
     $t_{1\alpha 1\beta 3\alpha 2\beta}$ & -0.00031383 & -0.00052074& -0.00085024 & 0.00259242\\
     $t_{1\alpha 1\beta 3\alpha 3\beta}$ & -0.0294424 & -0.03458878 & -0.0117965 & 0.00736854\\
     $t_{1\alpha 1\beta 3\alpha 4\beta}$ & -0.00046086 & 0.000494& 0.00071524 & 0.00102053\\
     $t_{1\alpha 1\beta 4\alpha 2\beta}$ &  -0.01874083 &  -0.04455011 &  -0.13145439 &  -0.14748198\\
     $t_{1\alpha 1\beta 4\alpha 3\beta}$ & -0.00016563 & 0.00001072 & 0.00024277 & -0.00042897\\
     $t_{1\alpha 1\beta 4\alpha 4\beta}$ & -0.0311721 & -0.05625665 & -0.03546193 & -0.02136681\\
     \hline
    \end{tabular}
\end{table*}

\end{document}